\documentclass[12]{article}
\usepackage[utf8]{inputenc}
\usepackage{epstopdf}
\usepackage{subfigure}
\usepackage{braket}

\usepackage{geometry}
\usepackage{booktabs}
\usepackage{multirow}
\usepackage{titlesec}
\usepackage[flushleft]{threeparttable}
\usepackage{xcolor}
\usepackage[bottom]{footmisc}
\usepackage{natbib}
\usepackage{graphicx}
\usepackage{graphics}
\usepackage{amsmath}
\usepackage{mathtools,amsfonts,amssymb,amsthm}
\usepackage{authblk}
\usepackage{appendix}

\usepackage{sectsty}
\sectionfont{\large}
\subsectionfont{\itshape}

\title{Mesoscopic Structure of the Stock Market\\and Portfolio Optimization}
\author[1]{S.M. Zema}
\author[1]{G. Fagiolo}
\author[2]{T. Squartini}
\author[2, 3]{D. Garlaschelli}
\affil[1]{Istituto di Economia, Scuola Superiore Sant'Anna, Pisa, IT}
\affil[2]{Networks, IMT Institute for Advanced Studies, Lucca, IT}
\affil[3]{Lorentz Institute for Theoretical Physics, University of Leiden, Leiden, NL}

\date{}

\begin{document}

\maketitle

\begin{abstract}
The idiosyncratic (microscopic) and systemic (macroscopic) components of market structure have been shown to be responsible for the departure of the optimal mean-variance allocation from the heuristic `equally-weighted' portfolio. In this paper, we exploit clustering techniques derived from Random Matrix Theory (RMT) to study a third, intermediate (mesoscopic) market structure that turns out to be the most stable over time and provides important practical insights from a portfolio management perspective. First, we illustrate the benefits, in terms of predicted and realized risk profiles, of constructing portfolios by filtering out both random and systemic co-movements from the correlation matrix. Second, we redefine the portfolio optimization problem in terms of stock clusters that emerge after filtering. Finally, we propose a new wealth allocation scheme that attaches equal importance to stocks belonging to the same community and show that it further increases the reliability of the constructed portfolios. Results are robust across different time spans, cross-sectional dimensions and set of constraints defining the optimization problem
\end{abstract}
\vspace{10pt}

\vspace{5pt}

\section{Introduction}

The pioneering work of \cite{Markowitz1952} laid the foundations of modern portfolio theory through the mean-variance (MV) optimization procedure. According to that model, the portfolio optimizer deals with uncertainty either by minimizing the variance of the investment, given the expected return, or by maximizing the expected return, given a certain level of risk. Despite its simplicity, it is widely recognized that the mean-variance framework delivers a poor out-of-sample performance \citep{michaud1989, bai2009enhancement}. The MV predictions, in fact, seriously depart from empirical observations, thus questioning the need of individuating procedures for investment optimization - especially when considering that the simpler $1/N$ heuristic rule, being less affected by covariance estimation errors, achieves better out-of-sample performances \citep{duchin2009}. Indeed, as \cite{laloux1999noise} show, the MV optimization procedure provides a biased estimation of the correlation matrix since the smallest eigenvalues of the latter, which play a fundamental role in the estimation of the global minimum variance (GMV) portfolio, are largely affected by noise.

In order to overcome the limitations of the MV framework, techniques have been proposed either to ameliorate its theoretical predictions or to capture the `real' essence of the correlation matrix by means of which optimal portfolios are constructed through filtering procedures\footnote{A comprehensive empirical study, regarding the possible improvements in the optimal asset allocation through the replacement of the sample correlations estimator with other estimation and filtering techniques, can be found in \cite{pantaleo2011improved}.}. Overall, each different estimator and filtering procedure improves upon different portfolio aspects related to the performance, realized risk, reliability and diversification and these improvements also depend on other circumstances as the dimension-to-sample size ratio or the possibility of exploiting short-selling strategies.

One network-based approach exploits the complex, evolving and interconnected nature of markets. For example, \cite{Onnela2003} and \cite{peralta2016network} point out the existence of a relationship between the centrality of each stock in the network of log-return correlations and the weight induced by the MV optimization procedure, an evidence suggesting that optimal portfolios should include peripheral stocks to reduce the influence of central assets characterized by higher levels of variance. Other studies heavily rely on hierarchical-clustering techniques \citep{mantegna1999hierarchical,bonanno2003topology,bonanno2004networks,di2004interest,onnela2004clustering,tumminello2005tool}: for instance, in \cite{TOLA2008235} optimal portfolios are constructed by replacing the empirical correlations with the ultrametric distances induced by the corresponding hierarchical-clustering scheme.

A different stream of literature focuses instead on filtering procedures that rely upon Random Matrix Theory (RMT) \citep{biely2008random,dimov2012hidden,singh2016random,zitelli2020random}. As shown in \cite{MacMahon}, different components of the market structure can be identified by employing an RMT-based clustering technique that returns cohesive groups of stocks on the basis of which the portfolio optimization problem can be reformulated. Such an approach has been recently adopted by \cite{anagnostou2021uncovering}, who have focused on Credit Default Swap (CDS) markets, showing that such structures are indeed useful for credit risk modelling, especially because they may encode factors not necessarily related with standard industry/region taxonomies. Taken together, these results point out that filtered correlation matrices are typically more reliable - in terms of predicted and realized risk profiles - than those obtained using the empirical correlations as input\footnote[3]{This is true especially when the requirement $T\gg N$ cannot be satisfied \citep{laloux2000random,plerou2002random}.}.

Moving from there, our work makes a step forward and investigates the effects of adjusting the correlation matrix of financial assets by considering not only the amount of correlations induced by noise but also the one induced by systemic co-movements. As documented by \cite{Forbes}, the correlation coefficient is indeed conditional on market volatility\footnote[4]{In the rest of the article we will refer to the terms \textit{systemic effects} and \textit{market effects} as synonyms.}, a direct consequence of which being that variables might appear as strongly correlated only because of temporary turmoil periods. For this reason, focusing on stable interconnections between stocks is fundamental to improve a portfolio reliability for the wealth allocation process: such a goal can be achieved by identifying which market correlations are structural and stable over time, i.e. not resulting from either random co-movements or temporary market effects.

As we will show, optimizing portfolios using the intermediate, mesoscopic level of the spectrum of the correlation matrix yields balanced allocations that improve the reliability of the former ones, in terms of predicted and realized risk, as compared to the standard MV optimization procedure: such an asset allocation closely tracks the heuristic $1/N$ rule but is not sensitive to estimation errors, caused either by random or aggregate systemic fluctuations, affecting correlations; this, in turn, reduces the effective size of portfolios without hampering their performance.

Furthermore, we show that redefining the asset allocation problem by giving equal importance to assets belonging to the same communities, i.e. groups of strongly interconnected stocks identified after filtering out noise and common aggregate effects \citep{MacMahon, anagnostou2021uncovering}, one is able to construct portfolios that are more reliable than those obtained by both the classical MV plug-in estimator\footnote[5]{In what follows, we refer to the historical plug-in estimator for the MV optimization as the `classical Markowitz' approach.} and the $1/N$ rule.

The rest of the paper is organized as follows. In section 2 we introduce our filtering procedure and explain how filtered correlations can be exploited to recover the mesoscopic structure of the stock market. In section 3 we show how that information can be exploited in a portfolio optimization setting. Section 4 illustrates the advantages, in terms of predicted and realized risk reliability, of constructing portfolios as above. Section 5 concludes and discusses possible paths for future research.

\section{The mesoscopic structure of the stock market}

RMT can be employed to filter out the random noise from the correlation matrices of financial returns, by exploiting the \textit{Mar\v{c}enko-Pastur Law} \citep{Marcenko1967}. More formally, let \{$x_{it}$\}, with $i=1\dots N$ and $t=1\dots T$, be a sample of i.i.d. random variables with zero mean and variance $\sigma^{2}$. Let $\kappa$ be the ratio $T/N$, assuming $\kappa\in (1,\infty)$ in the limit $T,N \rightarrow \infty$. Then, with probability one, the spectral density function of the sample covariance matrix tends to the Mar\v{c}enko-Pastur distribution, i.e.

\begin{equation}
f_{\kappa}(\lambda) = \frac{\kappa}{2\pi\lambda\sigma^{2}}\sqrt{(\lambda_\textrm{max} - \lambda)(\lambda - \lambda_\textrm{min})}  
\end{equation}
for $\lambda_\textrm{min}\leq\lambda\leq\lambda_\textrm{max}$, where $\lambda_\textrm{max}=\sigma^{2}(1 + \sqrt{N/T})^{2}$ and $\lambda_\textrm{min}=\sigma^{2}(1 - \sqrt{N/T})^{2}>0$.

\noindent The reader interested in the proof is redirected to \cite{Bai1999}. The result above implies that any empirical correlation matrix $\mathbf{C}$ of financial returns\footnote[6]{Notice that the correlation matrix coincides with the covariance matrix of standardized returns: in this case, $\sigma^{2}=1$ and the range extremes simply read $\lambda_\textrm{min/max}= (1 \pm \sqrt{N/T})^{2}$.}, where the largest empirical eigenvalue is denoted as $\lambda_m$ and is usually (much) larger than $\lambda_\textrm{max}$, can be decomposed as

\begin{eqnarray}
\mathbf{C}& = &\sum_{i=1}^N\lambda_i|v_i\rangle\langle v_i|\nonumber\\
&=&\sum_{i:\lambda_i\in(0,\lambda_\textrm{max}]}\lambda_i|v_i\rangle\langle v_i|+\sum_{i:\lambda_i\in(\lambda_\textrm{max},\lambda_m]}\lambda_i|v_i\rangle\langle v_i|\\
&=&\mathbf{C}^{(r)}+\mathbf{C}^{(s)},
\end{eqnarray}
where $|v_i\rangle$ and $\langle v_i|$ denote the column and row eigenvectors associated with the eigenvalue $\lambda_i$ respectively. The above decomposition represents the empirical correlation matrix as a sum of matrices respectively induced by the \emph{random} spectral component $\mathbf{C}^{(r)}$, whose eigenvalues lie in the Mar\v{c}enko-Pastur range $[\lambda_\textrm{min},\lambda_\textrm{max}]$ and usually also below $\lambda_\textrm{min}$ (as a result of the fact that, since the trace of a correlation matrix should remain equal to $N$, the presence of $\lambda_m>\lambda_\textrm{max}$ shifts the lower eigenvalues leftwards) and the \emph{structural} (non-random) component $\mathbf{C}^{(s)}$. The filtering procedure consists in removing $\mathbf{C}^{(r)}$, i.e. the random component of the tensor: what remains is, then, recognized as signal rather than noise, hence supposed to possess useful economic information.

As mentioned above, the spectrum of empirical correlation matrices of financial returns is characterized by a leading eigenvalue $\lambda_m$ which is much larger than the others. The associated (column) eigenvector $|v_m\rangle$ possesses elements having the same sign and identifies a matrix component $\lambda_m|v_m\rangle\langle v_m|$ affecting all stocks in the same direction and with strong intensity, further inducing the decomposition of $\mathbf{C}^{(s)}$ as follows:

\begin{equation}
\mathbf{C}^{(s)}=\sum_{i:\lambda_i\in(\lambda_\textrm{max},\lambda_m)}\lambda_i|v_i\rangle\langle v_i|+\lambda_m|v_m\rangle\langle v_m|=\mathbf{C}^{(g)}+\mathbf{C}^{(m)}
\end{equation}
i.e. as a sum of a \emph{mesoscopic} spectral component $\mathbf{C}^{(g)}$ and a \emph{systemic} component (or \emph{market mode}) $\mathbf{C}^{(m)}$. 

\begin{figure}
\begin{center}
\includegraphics[width=\textwidth]{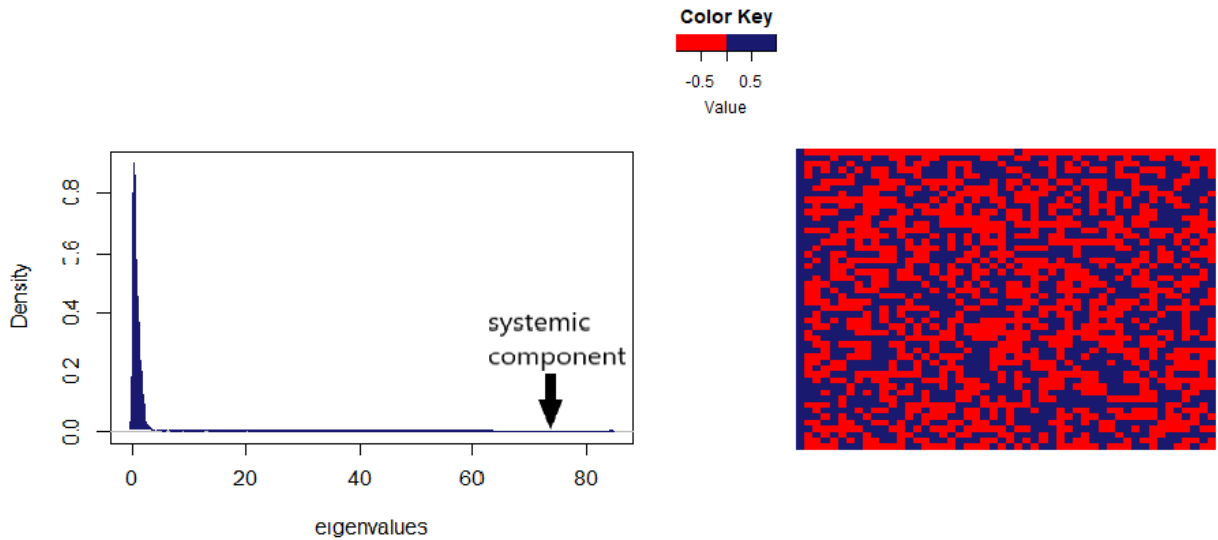}
\caption{Eigenvalue density for the 2000-2015 covariance matrix of the S\&P500 components (left) and heatmap of the associated eigenvectors (right). The first column of the heatmap represents the eigenvector associated to the systemic component, whose elements all have the same sign.}\label{1}
\end{center}
\end{figure}

A graphical illustration of this empirical feature is provided in Figure \ref{1} for the S\&P500 constituents over the period 2000-2015 which constitute the dataset used in the remainder of the work as well. The systemic component is pervasive and time-varying; hence, some stocks might appear interconnected only as a consequence of their common dependence on global market events \citep[see][among others]{Forbes,billio2012econometric}. When performing asset allocation strategies based on historical data, it is fundamental to minimize covariance estimation errors induced by any possible time-varying component - which makes historical data not reliable for the future.

\begin{figure}
\begin{center}
\includegraphics[width=0.75\textwidth]{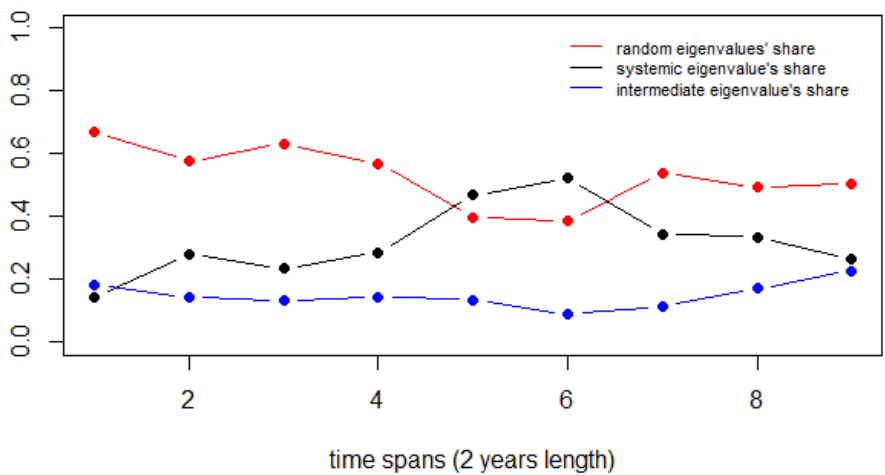}
\caption{Cumulative risk fractions associated to the different components of the correlation matrix over different time spans. The random and systemic components vary the most (14\% standard deviation for both) while the intermediate, mesoscopic range of the spectrum is more stable (5\% standard deviation only)}\label{2}
\end{center}
\end{figure}

To provide an example, let us define the total risk of a system as $\Lambda :=\sum_{k=1}^{N}\lambda_{k}$ and investigate the temporal evolution of the cumulative risk fraction of the different components of the covariance matrix of stock returns by adopting non-overlapping, rolling windows of two years. To this aim, can draw 100 randomized samples of size 100 from the S\&P500 constituents, for each temporal window: the resulting averaged shares of total risk accounted by the random, systemic and mesoscopic component of the spectrum of the covariance matrix are shown in Figure \ref{2}. While the random and systemic cumulative risk fractions vary quite a lot across the considered period, the mesoscopic one is the most stable, a result letting us to conclude that the construction of more reliable portfolios may indeed be based on the stable part of the spectrum.\\

Let us now use $\mathbf{C}$ to partition the stock market into non-overlapping communities of stocks that are more correlated internally than expected under a suitable null model. 
Detecting communities in financial markets is not new in the literature: for instance, \cite{fenn2012dynamical} compared different procedures to unfold the community structure of the foreign exchange market and \cite{verma2019cluster} used clusters to extract relevant factors for volatility modeling. However, the procedure we are now going to illustrate is based on a combination of modularity maximization \citep{clauset2004finding,Newman8577} and RMT \citep{MacMahon}, which was shown to be theoretically superior in the case of correlation matrices. 

In the network science literature, the so-called modularity $Q(\boldsymbol{\gamma})$ of a partition $\boldsymbol{\gamma}$ of the $N$ nodes of a network is defined as

\begin{equation}
Q(\boldsymbol{\gamma}) =\frac{1}{\sum_{i=1}^N\sum_{j=1}^Nw_{ij}}\sum_{i=1}^N\sum_{j=1}^N\left[w_{ij}- \langle w_{ij} \rangle \right]\delta(\gamma_{i},\gamma_{j})
\end{equation}\\
where $w_{ij}$ the entry of the adjacency matrix of the (possibly weighted) network (i.e. $w_{ij}$ is the weight of the link from node $i$ to node $j$), $\langle w_{ij} \rangle$ is its expected value under a suitably chosen null model, and the Kronecker delta $\delta(\gamma_{i},\gamma_{j})$ guarantees that only the nodes belonging to the same community contribute to the modularity. The goal of modularity maximization is finding the partition that maximizes $Q(\boldsymbol{\gamma})$, thus emphasizing the community of nodes whose internal interactions are stronger and maximally unexplained by the (community-free) null model.

For networks, the null model chosen is generally the so-called Weighted Configuration Model (WCM) that randomizes the network topology while preserving the empirical strength $s_{i}=\sum_{j=1}^{N}w_{ij}$ of each node $i$. A popular, although in general incorrect \citep{garlaschelli2009generalized}, expression used to represent this null model is

\begin{equation}
\langle w_{ij}\rangle=\frac{s_{i}s_{j}}{2W}\quad\forall\:i,j
\end{equation}
where $2W=\sum_{i=1}^{N}s_{i}=\sum_{i=1}^N\sum_{j=1}^Nw_{ij}$ is the total edge weight of the network. When considering correlation matrices, the null model above has been shown to be inconsistent \citep{MacMahon} as a result of the fact that, unlike (weighted) networks, correlation matrices cannot be directly randomized by considering their entries as independent. Rather, the randomization should occur at the level of the underlying time series, and the correlation matrix should then be recalculated from the randomized time series. In particular, by reformulating the modularity for correlation matrices as

\begin{equation}
Q(\boldsymbol{\gamma}) =\frac{1}{\sum_{i=1}^N\sum_{j=1}^NC_{ij}}\sum_{i=1}^N\sum_{j=1}^N\left[C_{ij}- \langle C_{ij} \rangle \right]\delta(\gamma_{i},\gamma_{j}),
\end{equation}
\noindent a consistent community-free null model representing random empirical correlations resulting only from noise and possibly global trends comes precisely from RMT and can be expressed as

\begin{equation}
\langle C_{ij}\rangle=C_{ij}^{(r)}+C_{ij}^{(m)}
\end{equation}
\citep{MacMahon}. The above null model discounts both the random and the systemic components of correlations. As a consequence, 

\begin{equation}
C_{ij}- \langle C_{ij} \rangle = C_{ij}^{(g)},
\end{equation}

\noindent i.e. the modularity matrix coincides with the mesoscopic component of the original correlation matrix. Therefore maximizing the modularity $Q(\gamma)$ guarantees that the identified communities are necessarily formed by internally positively (after discounting the null model) and mutually negatively (again, after discounting the null model) correlated stocks. In other words, the communities are ideally noise-free and mutually anti-correlated with respect to the market.

\begin{figure}
\begin{center}
\includegraphics[width=\textwidth]{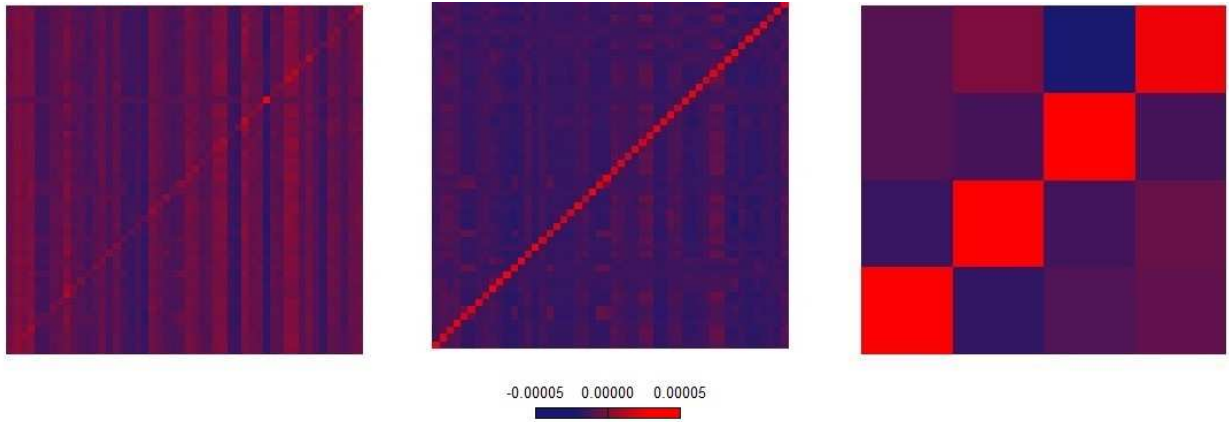}
\caption{\footnotesize{Sequence of transformations applied to the empirical correlation matrix (left), leading first to the `noise and systemic free' correlation matrix (middle) and then to the internally positively and mutually negatively correlated clusters.}}\label{3}
\end{center}
\end{figure}

\section{Stock market communities}

The dataset used for the present analysis has been downloaded from Yahoo Finance and consists of equity data for the 450 most capitalized companies in the US, stably traded over the last 20 years, all constituting the S\&P500.
After applying RMT to isolate the mesoscopic component of the matrix, we performed the modularity maximization procedure by implementing a modified version of the Louvain algorithm \citep{blondel2008fast}, taking as input the matrix $\mathbf{C}^{(g)}$. The consistency and stability of this approach have been discussed in \cite{MacMahon} and \cite{anagnostou2021uncovering} to which the interested reader is referred for additional technical clarifications. 

\begin{figure}
\begin{center}
\includegraphics[width=0.75\textwidth]{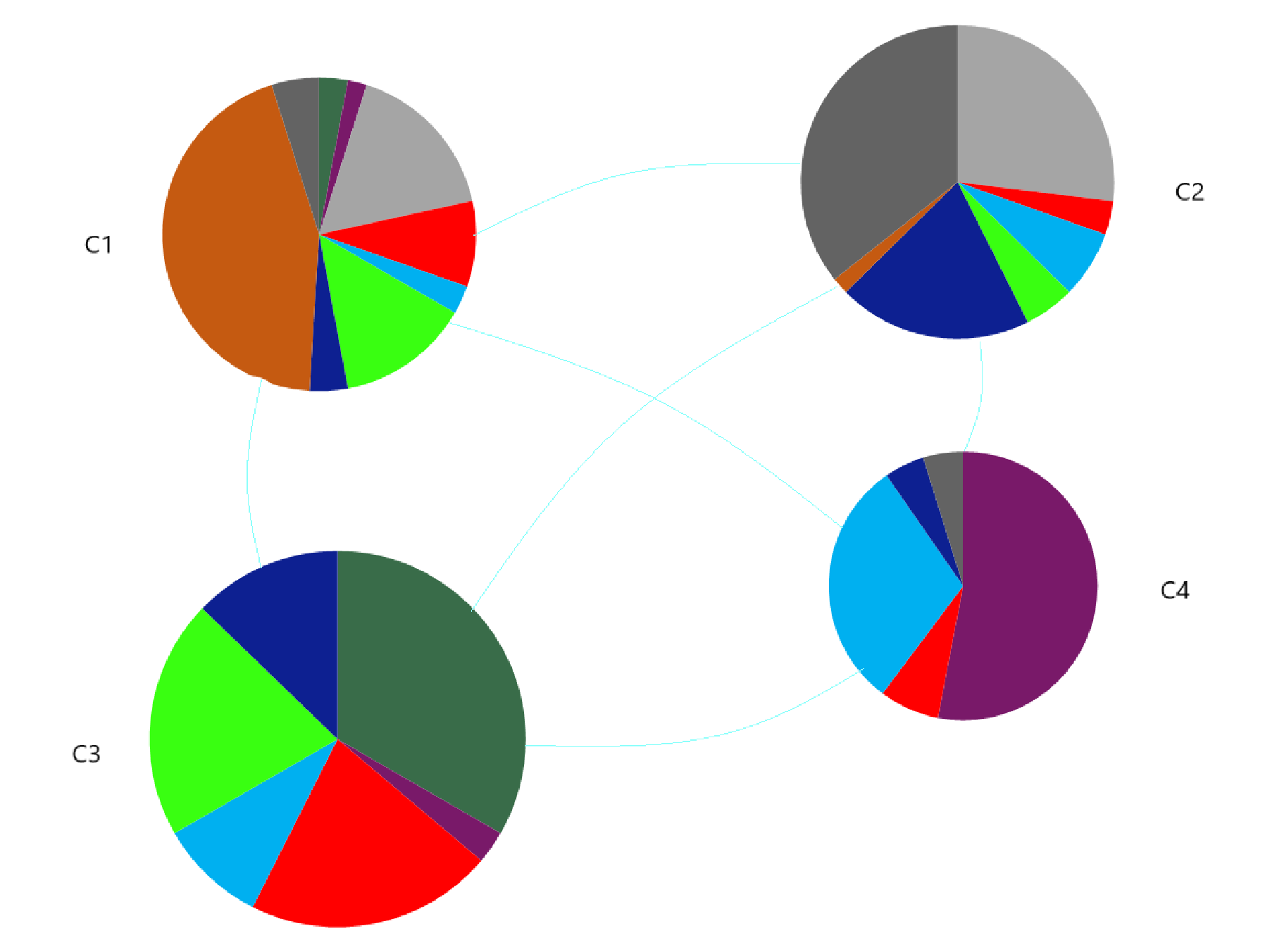}
\caption{\footnotesize{Community structure of the 450 most capitalized stocks of the US stock market during the period January 2000-December 2015: $\textcolor{teal}{\blacksquare}$ Finance, $\textcolor{violet}{\blacksquare}$ Energy, $\textcolor{gray!70}{\blacksquare}$ Healthcare, $\textcolor{red}{\blacksquare}$ Industrials, $\textcolor{cyan}{\blacksquare}$ Materials, $\textcolor{green}{\blacksquare}$ Discretionary, $\textcolor{blue}{\blacksquare}$ Staples, $\textcolor{brown}{\blacksquare}$ Technology, $\textcolor{darkgray}{\blacksquare}$ Utilities.}}\label{4}
\end{center}
\end{figure}

We conducted the above analysis across the time span 2000-2015 and identified an optimal partition of the 450 stocks into 4 communities.
Figure \ref{3} shows the heatmaps depicting the sequence of transformations leading from the original stocks to such a set of mutually, negatively correlated communities. Figure \ref{4} shows the detected communities, together their relative compositions, according to the industrial classification. The number of detected communities is lower than the number of considered sectors, showing the tendency of stocks to be strongly interconnected across different sectors as well. Still, it can be noticed how stocks belonging to specific sectors tend to cluster more than others - a behavior detected also in \cite{Borghesi2007} by employing hierarchical clustering techniques. In particular, almost all stocks in the financial sector are clustered together in C3 while stocks in the energetic and technological sectors are respectively placed in C4 and C1; utilities are quite clustered in C2. The remaining sectors (namely the industrials, materials, consumer discretionary, consumer staples and healthcare), instead, are more dispersed across different communities. This result confirms that the data-driven cluster identification leads to communities that are unpredictable from the nominal sectoral classification of stocks, as also observed in \citet{MacMahon} and \citet{anagnostou2021uncovering}.\\

\begin{figure}
\begin{center}
\includegraphics[width=0.75\textwidth]{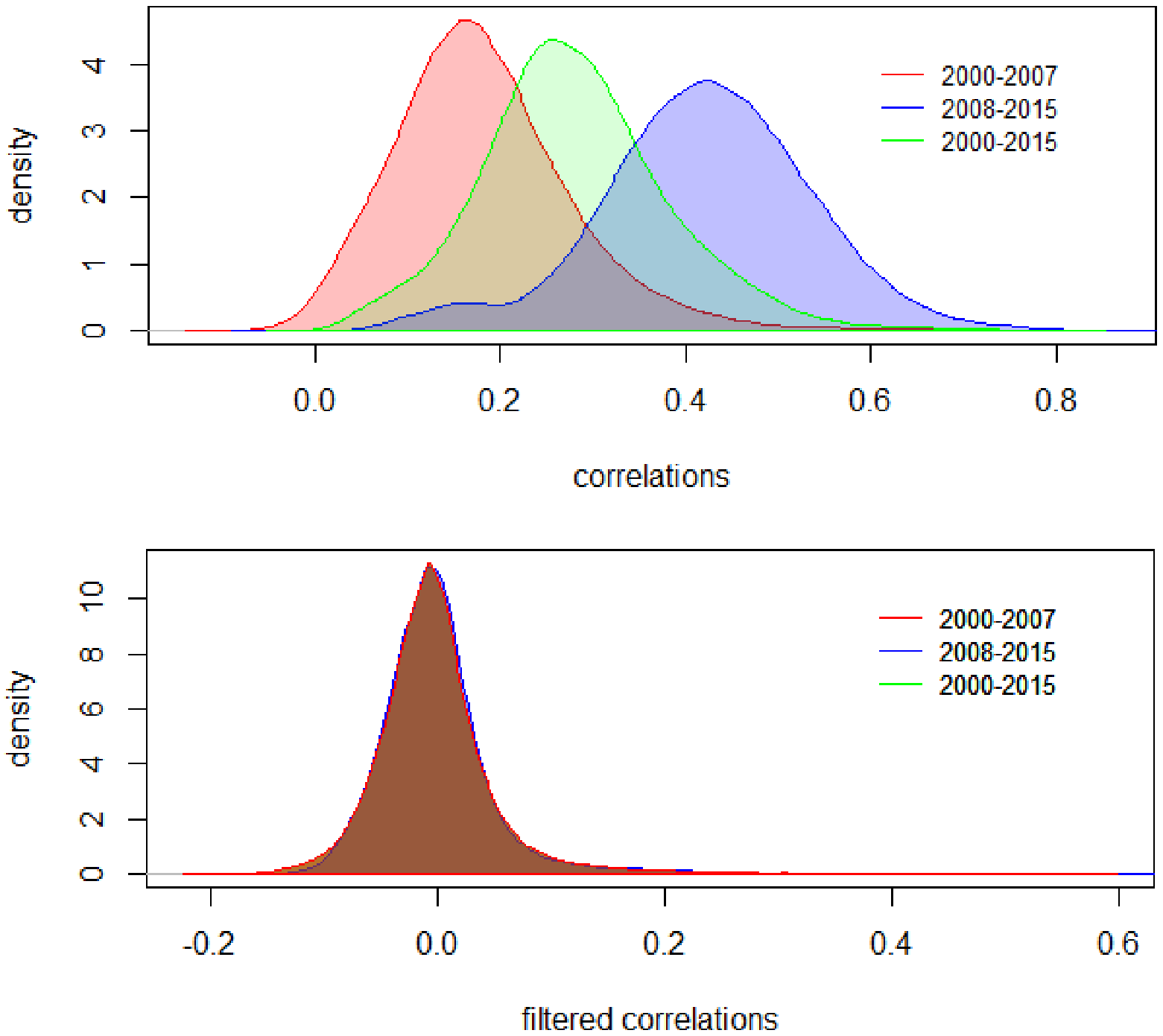}
\caption{\footnotesize{Densities of the unfiltered empirical (top) and filtered mesoscopic (bottom) correlation coefficients for S\&P constituents over different periods. Notice that, while a clear shift occurs between the first and the second half of the overall 2000-2015 period for the unfiltered coefficients, no shift occurs for the filtered mesoscopic ones. As a consequence, the distribution of the unfiltered matrix entries calculated over the entire period is not representative of the distributions for the individual sub-periods, while that of the filtered matrix entries is.}}\label{5}
\end{center}
\end{figure}

Performing the community detection over the whole available time span, i.e. from 2000 to 2015, might seem unreasonable since structural changes have arguably occurred in occasion of the 2008 financial crisis and possibly other events. This is only partially true: it turns out that while the original, unfiltered empirical correlations do change a lot over time (especially during market turmoils), mesoscopic correlations remain remarkably stable, in turn stabilizing the optimal partition. This can be easily seen by comparing the evolution of the density of the unfiltered, empirical correlations of our sample of stocks with that of the filtered, mesoscopic correlations employed to perform the clustering procedure. As Figure \ref{5} shows, the distribution of the empirical correlation coefficients clearly shifts toward higher values in the second half of the considered time span (which contains a period of higher turmoil), so that the coefficients calculated over the entire time span are not representative of the underlying sub-periods. 
By contrast, when considering only the mesoscopic component of the correlation matrix over different periods, we find that the distribution of the entries of such component almost perfectly overlap with each other over time. In this case, the overall distribution is representative of the distributions for the sub-periods.

Since the stability of a distribution does not necessarily imply the stability of the individual entries of the matrix components, as a more stringent test we look at the evolution of the norm of the matrix containing the relative changes of each coefficient. Given the generic entry $C_{ij}^{(f)}$, where $f$ indicates which matrix component we are considering, we define the matrix $\Delta\mathbf{ C}^{(f)}$ with entries

\begin{equation}
\Delta C_{ij}^{(f)}=\frac{C_{ij}^{(f)}(t)-C_{ij}^{(f)}(t-1)}{C_{ij}^{(f)}(t-1)}.
\end{equation}
\noindent The above quantity is the relative variation of the correlation coefficient across two consecutive time spans. We focus on relative variations since the components of the correlation matrix are characterized by magnitudes that are so different that a comparison of absolute changes would be meaningless. The stability of each component of the correlation matrix can be inspected by employing the $p$-norm $||\Delta \mathbf{C}^{(f)}||_{p}$. As we can see from Table \ref{T1}, the mesoscopic component $\mathbf{C}^{(g)}$ turns out to be the most stable over time, given its smallest temporal variations.

To further assess the stability of the mesoscopic part of the correlation matrix, we repeat the computation just described by arranging the terms of $\mathbf{C}^{(g)}$ according to the community a given node belongs to, so to have a matrix arranged in blocks, as displayed in Figure \ref{3}. Afterwards, to attach equal importance to nodes according to the community they belong to, we replace the values in each given block with the average computed over the block itself and denote this new matrix as $\mathbf{C}^{(\overline{g})}$. Notice that no clear difference arises with respect to the norms computed for $\mathbf{C}^{(g)}$, a result confirming the stability of the detected structures.

Taken together, all the above checks of the stability of the filtered, mesoscopic component of empirical correlations lay the ground for our subsequent analyses in the rest of the paper.

\begin{table}[h]
\centering
\begin{threeparttable}
\begin{tabular}{c c c c c}
         & $|| \Delta \textbf{C}^{(r)} ||_{1}$ & $|| \Delta \textbf{C}^{(m)} ||_{1}$ & $|| \Delta \textbf{C}^{(g)} ||_{1}$ &$|| \Delta \textbf{C}^{(\overline{g})} ||_{1}$ \\
\hline 
 $\Delta T_{1}$  &    10.02            &         2.69               &    0.59     &   0.64   \\
 $\Delta T_{2}$  &    15.51            &         8.93               &    1.87     &   1.83  \\
 $\Delta T_{3}$  &    8.58             &         0.81               &    0.67     &   0.68   \\
\hline
\end{tabular}
\caption{Norms of the entry-by-entry relative changes of the correlation coefficients associated to the different components of the matrix. $T_1$, $T_2$ and $T_3$ denote time spans of equal length covering the period 2000-2015. The 1-norm has been chosen for simplicity.}\label{T1}
\end{threeparttable}
\end{table}

\section{Back to basic portfolio optimization}

Let us now address the implications of the market mesoscopic structure from a portfolio management perspective. In order to do so, let us briefly review the classical Markowitz portfolio optimization scheme. Consider $N$ risky assets with covariance matrix $\boldsymbol{\Sigma}$ and vector of expected returns $\boldsymbol{\mu}$. Given the wealth allocation vector $\boldsymbol{\omega}=[\omega_{1}\dots\omega_{N}]$, such that $\sum_{i}\omega_{i}=1$, the portfolio expected return reads

\begin{equation}
\mu_{p} = \sum_{i=1}^{N} \omega_{i}\mu_{i}
\end{equation}
with associated variance reading

\begin{equation}
\sigma_{p}^{2} = \sum_{i=1}^N \omega_{i}^2\sigma_{i}^2 + \sum_{i>j}2\omega_{i}\omega_{j}C_{ij}\sigma_{i}\sigma_{j}.
\end{equation}

The well-known Markowitz approach consists in finding the allocation vector $\boldsymbol{\omega}$ which minimizes $\sigma_{p}^{2}$ subject to a given value of $\mu_{p}$ or, equivalently, the one that maximize the return subject to a given level of variance. The optimization problem to be solved, expressed in matrix form, reads

\begin{equation}
\begin{aligned}
\min_{\boldsymbol{\omega}} \quad & \boldsymbol{\omega'}\boldsymbol{\Sigma}\boldsymbol{\omega}\\
\textrm{s.t.} \quad & \mu_{p} = \boldsymbol{\omega'}\boldsymbol{\mu}\\
 & \sum_{i}^{N}\omega_{i}=1
\end{aligned}
\end{equation}
and has solution

\begin{equation}
\boldsymbol{\omega^{*}} = \boldsymbol{b}\boldsymbol{\Sigma^{-1}}\boldsymbol{1} + \boldsymbol{c}\boldsymbol{\Sigma^{-1}}\boldsymbol{\mu}
\end{equation}
with

\begin{align*}
\boldsymbol{b} &= \frac{A -\boldsymbol{\mu_{p}}B}{\Delta} & \boldsymbol{c} &= \frac{\boldsymbol{\mu_{p}}C - B}{\Delta}\\
A &=\boldsymbol{\mu}'\boldsymbol{\Sigma^{-1}}\boldsymbol{\mu} & B &=\boldsymbol{1}'\boldsymbol{\Sigma^{-1}}\boldsymbol{\mu}\\
C &=\boldsymbol{1}'\boldsymbol{\Sigma^{-1}}\boldsymbol{1} & \Delta &= CA - B^{2}.
\end{align*}

In this work, we focus on the variance and consider a completely risk-adverse investor that is only interested in minimizing the risk with no constraints on the expected return. In that case, the solution simply becomes

\begin{equation}
w_{gmv} = \frac{\boldsymbol{\Sigma}^{-1}\boldsymbol{1}}{{\boldsymbol{1}}^{t}\boldsymbol{\Sigma}^{-1}\boldsymbol{1}}
\end{equation}
where $\omega_{gmv}$ denotes the investment plan associated with the global minimum variance (GMV) portfolio. 


We then focus on the \textit{reliability} of an optimal portfolio by comparing its predicted risk, $\sigma_{p}$, obtained via the correlation matrices estimated using historical data, with the (ex-post) realized risk, $\sigma_{p}^{r}$. As in \cite{TOLA2008235}, we deem a portfolio as reliable if

\begin{equation}
\mathcal{R}=\frac{|\sigma_{p}^{r}-\sigma_{p}|}{\sigma_{p}}
\end{equation}
is `small' - the main difference of our approach being that we will never assume perfect knowledge of future volatilities for the investor, letting uncertainty affect the whole covariance matrix.

\subsection{Noise-free and systemic-free optimization}

As shown before, the systemic component affects all stocks in the same direction, inducing a positive amount of covariance between the variables, i.e. $\sigma_{ij}^{(m)}>0 $, it is straightforward to show that, for a risk-minimizer investor, the adoption of the mesoscopic variance $\sigma_{ij}^{(g)} = C_{ij}^{(g)}\sigma_{i}\sigma_{j}$, in place of $C_{ij}^{(m)}$ holds) rebalances the portfolio: hence, the total wealth will not be concentrated anymore over few assets characterized by the lowest past variances. In other words, while in presence of co-movements (e.g. because of market turmoils), an `ingenuous' investor would (try to) lower the portfolio risk by concentrating the wealth over the less risky assets, an investor who is aware of the temporarily nature of aggregate shocks causing crashes in the market, would filter out the systemic effects from past data and trust only the stable part of the correlation matrix, for the future.

\begin{figure}
\begin{center}
\includegraphics[width=0.75\textwidth]{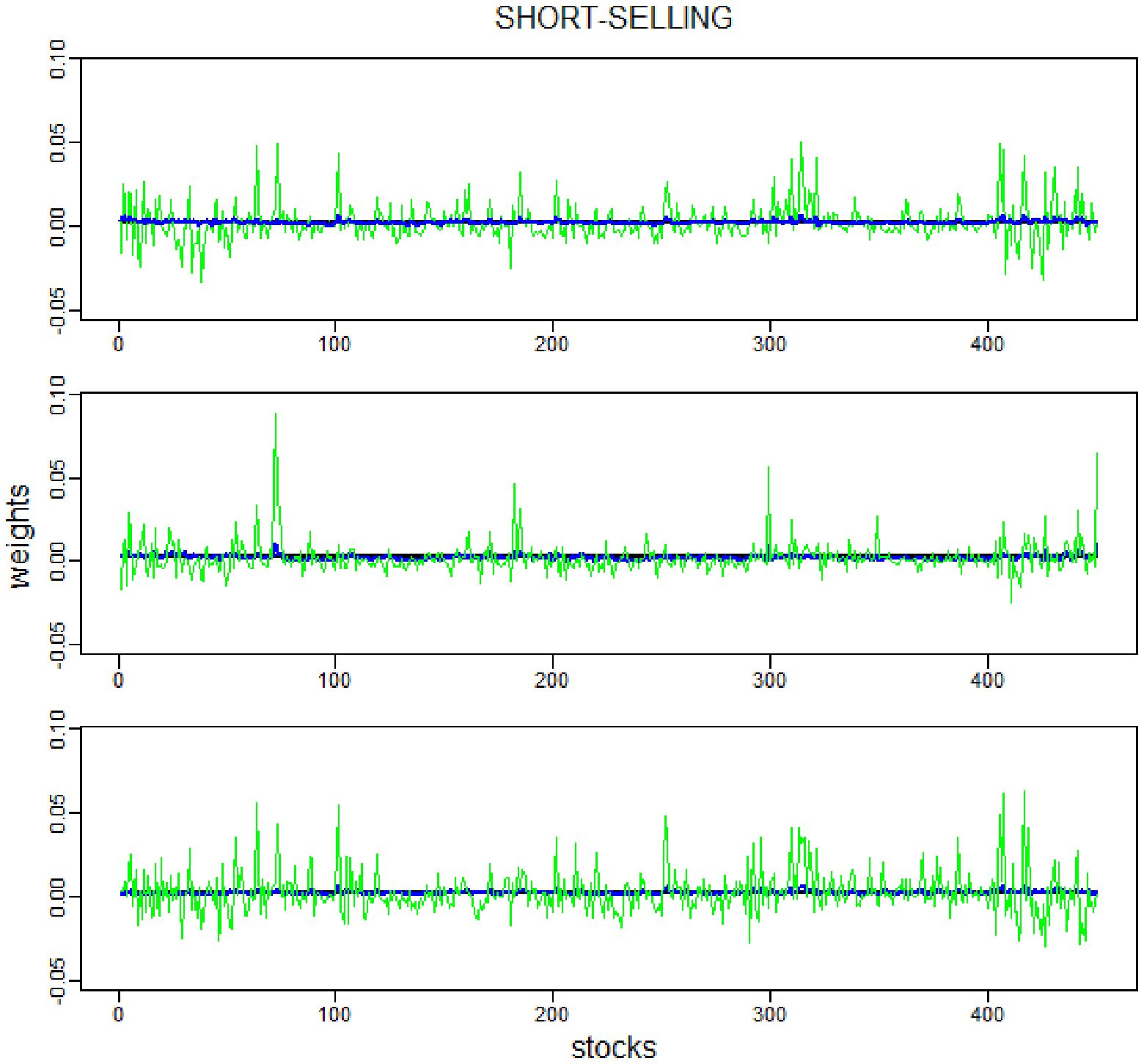}
\includegraphics[width=0.75\textwidth]{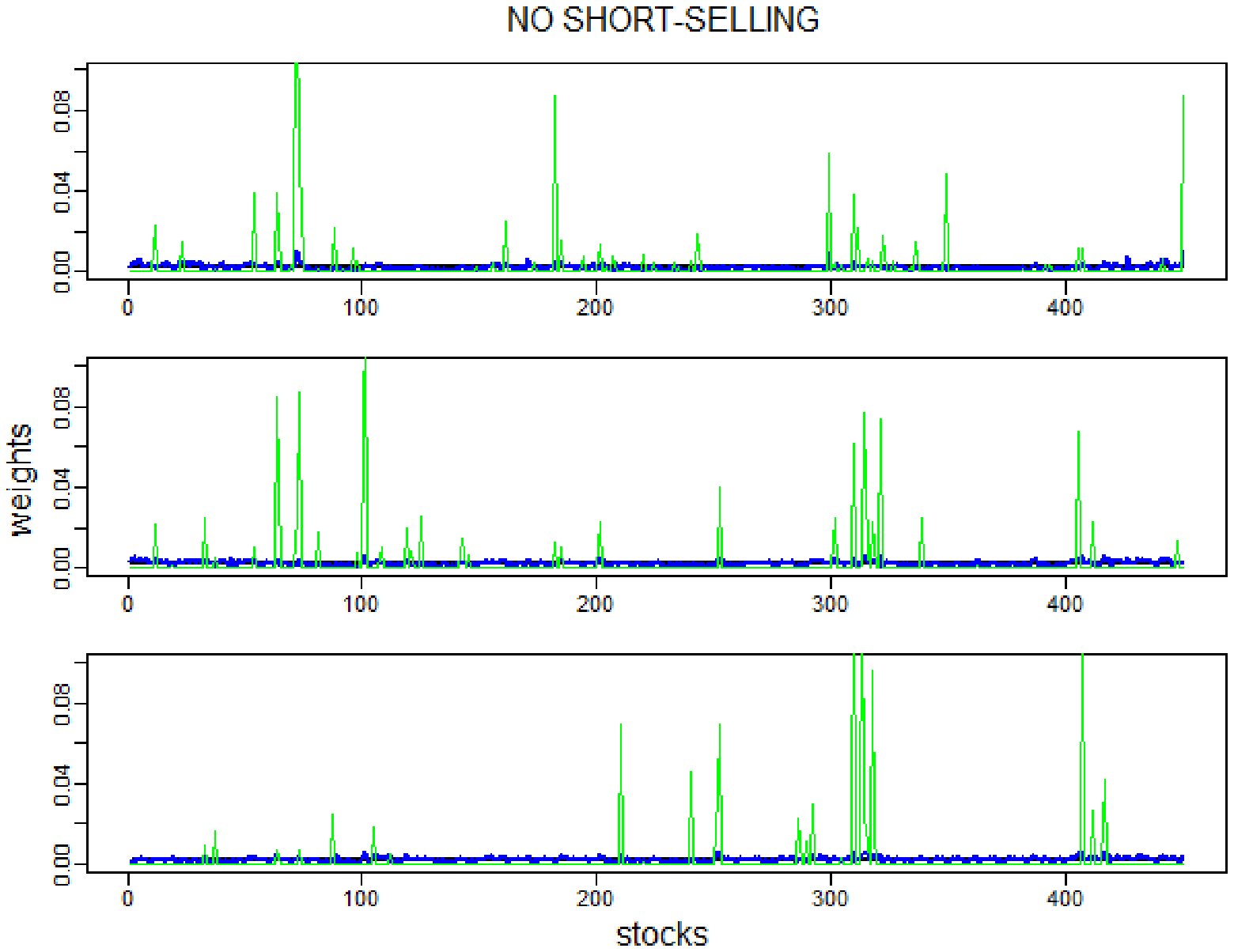}
\caption{\footnotesize{Asset composition comparison for the periods 2000-2003 (top), 2004-2007 (middle) and 2008-2011 (bottom)  between the $1/N$ rule (horizontal black line), RMT approach (green) and the portfolio optimization based on mesoscopic correlations (blue). For each stock on the x-axis, the relative weight on the y-axis is shown. Cleaning from the noise is not sufficient to closely track the heuristic rule as it is when adjusting from the market component as well.}}\label{7}
\end{center}
\end{figure}

To provide empirical evidence for such a statement, we performed the MV optimization procedure on the S\&P500 constituents over multiple periods, comparing the wealth allocation vectors obtained using the empirical and the mesoscopic correlation matrices and keeping the equally-weighted portfolio as a benchmark\footnote[7]{A short analytical description of the rebalancing effect for the $N=2$ assets case is provided in the appendix.}: Figure \ref{6} shows the results, considering both cases in which short selling is either possible or not. Noticeably, the MV optimization based on mesoscopic correlations closely follow the $1/N$ rule, yielding as an optimal solution a portfolio which is very similar to the equally-weighted one; on the contrary, the standard Markowitz optimization framework outputs a much more heterogeneous composition being more sensitive to the estimated sample covariances. This confirms what expected, i.e. that the optimization procedure based on the mesoscopic structure of the correlation matrix is less sensitive to both noisy and aggregated fluctuations, thus yielding more balanced portfolios. As an additional test, in Figure \ref{7} we compare the mesoscopic and $1/N$ weights with the ones we would obtain by cleaning the correlation matrix only from noise through the standard RMT-based approach: in order to closely track the balanced $1/N$ allocation it is necessary to filter out both the noisy and the systemic components.\\

\begin{figure}
\begin{center}
\includegraphics[width=0.75\textwidth]{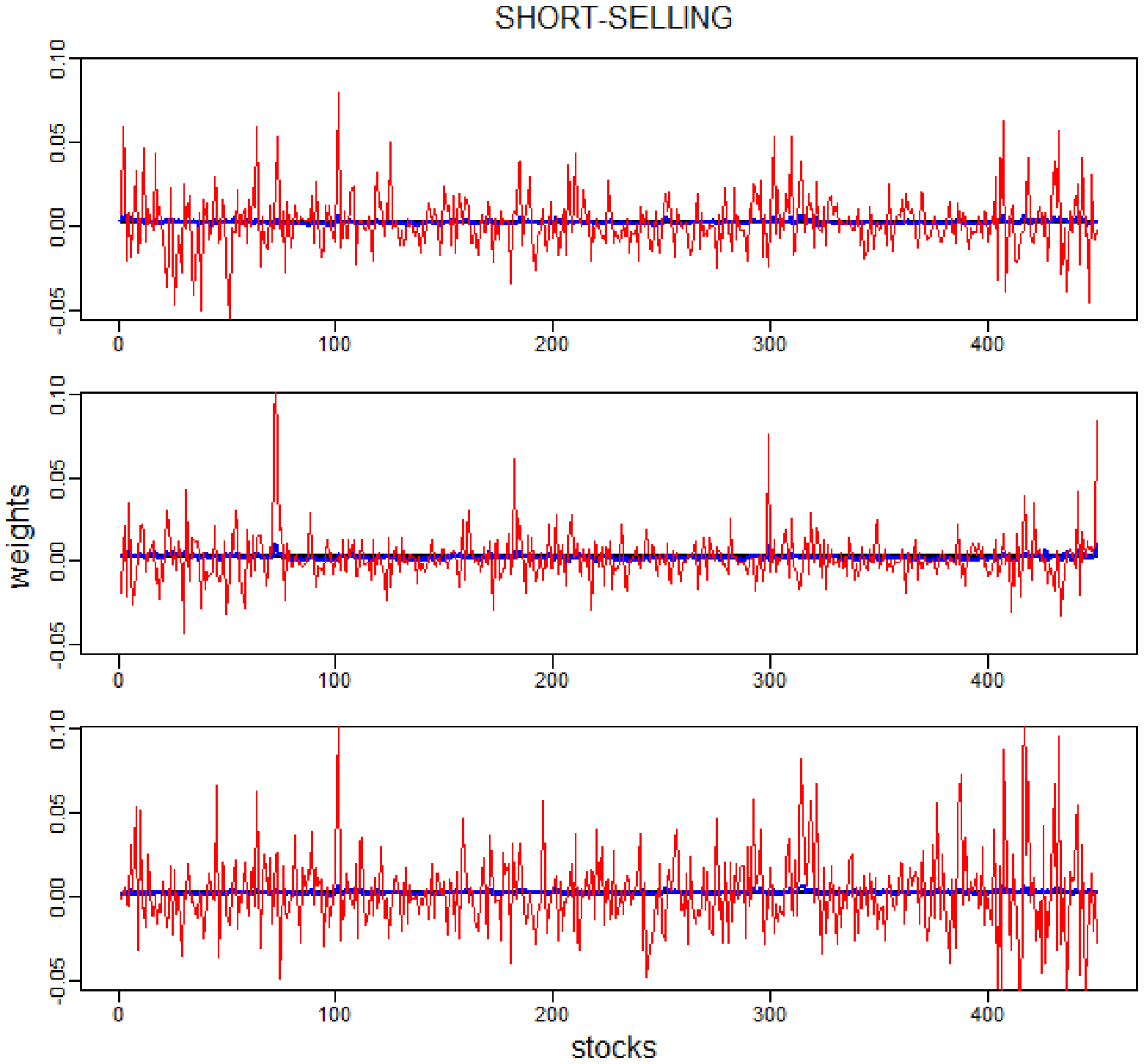}
\includegraphics[width=0.75\textwidth]{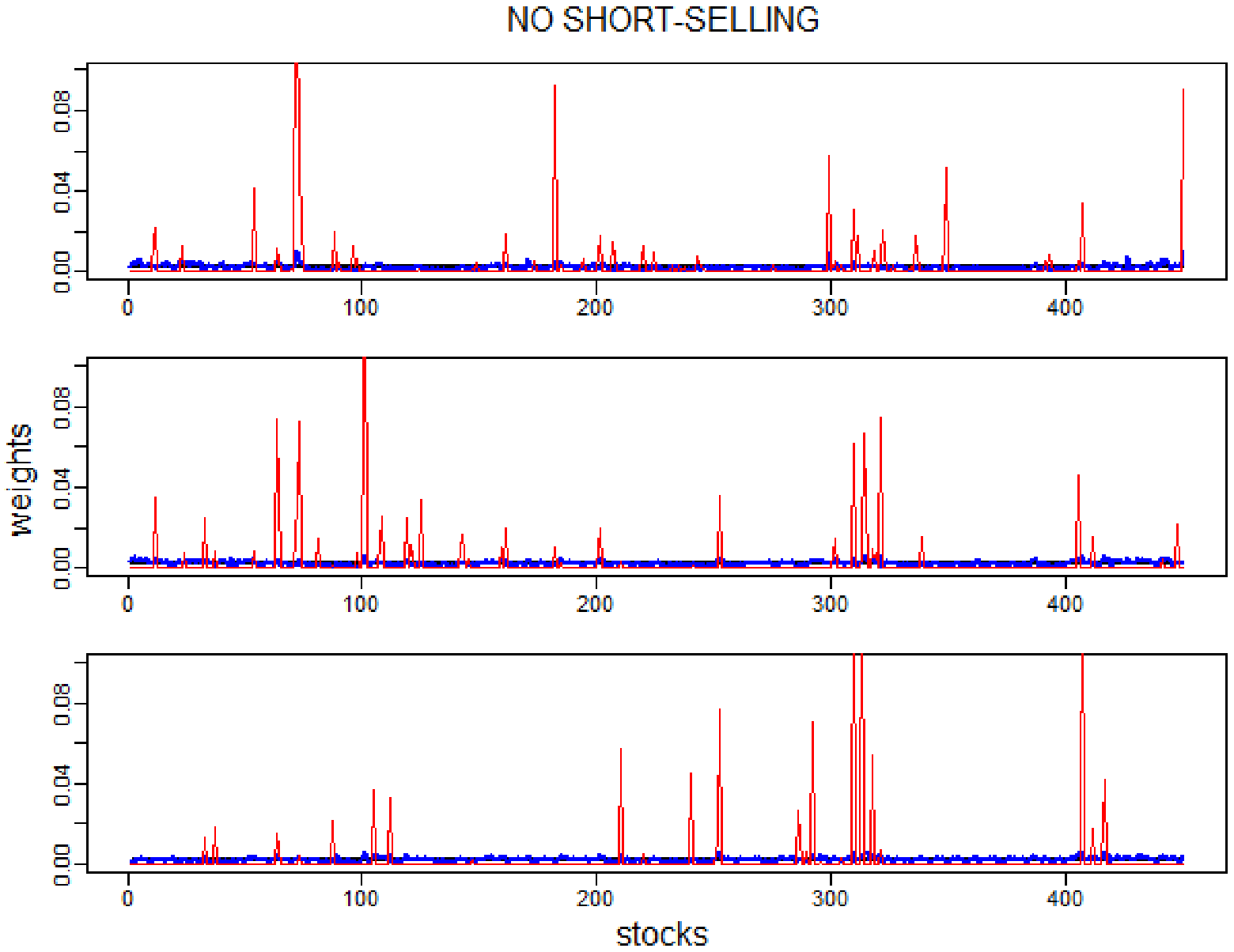}
\caption{\footnotesize{Asset composition comparison for the periods 2000-2003 (top), 2004-2007 (middle) and 2008-2011 (bottom)  between the $1/N$ rule (horizontal black line), classical Markowitz (red) and the portfolio optimization based on mesoscopic correlations (blue). For each stock on the x-axis, the relative weight on the y-axis is shown, the mesoscopic-based optimization closely follow the heuristic $1/N$ rule. When short-selling is not allowed we have $\omega_{i}\geq 0$, $\forall\:i$.}}\label{6}
\end{center}
\end{figure}

A measure of similarity to the equally-weighted portfolio is provided by the number of stocks with a `significant' amount of money invested into. Following \cite{bouchaud2003theory}, this quantity can be defined as

\begin{equation}
\mathcal{N} = \frac{1}{\sum_{i=1}^{N}\omega_{i}^{2}};
\end{equation}
indeed, when the wealth is equally divided among the $N$ assets, the quantity $\mathcal{N}$ is equal to $N$; on the other had, it is equal to 1 when the wealth is invested only in  one asset. As stressed in \cite{TOLA2008235}, the quantity $\mathcal{N}$ simply provides a rough estimate of the number of stocks which could be effectively used to build a portfolio that is smaller than the original but preserves most of the risk-return properties of the latter.

\begin{figure}
\begin{center}
\includegraphics[width=0.75\textwidth]{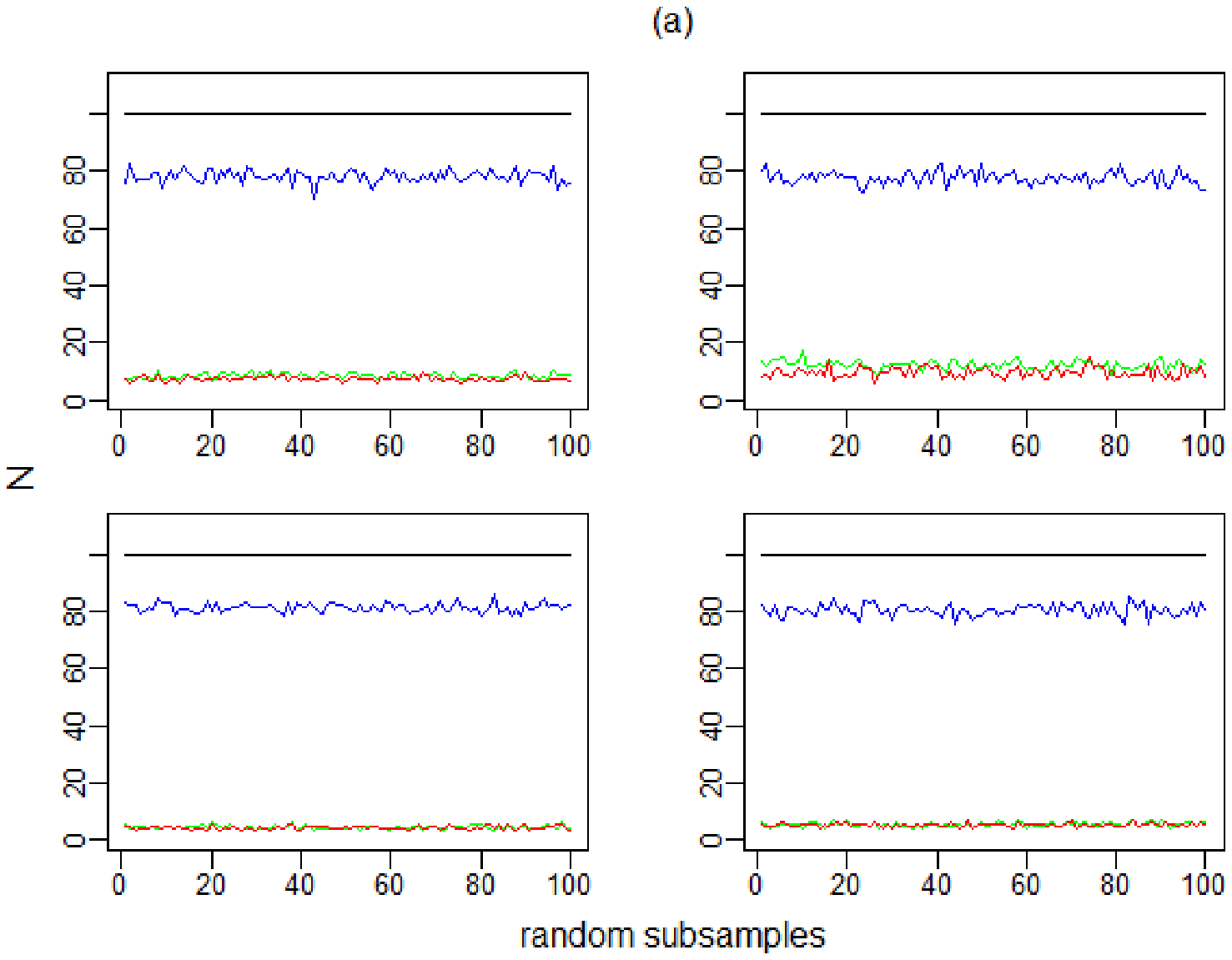}
\includegraphics[width=0.75\textwidth]{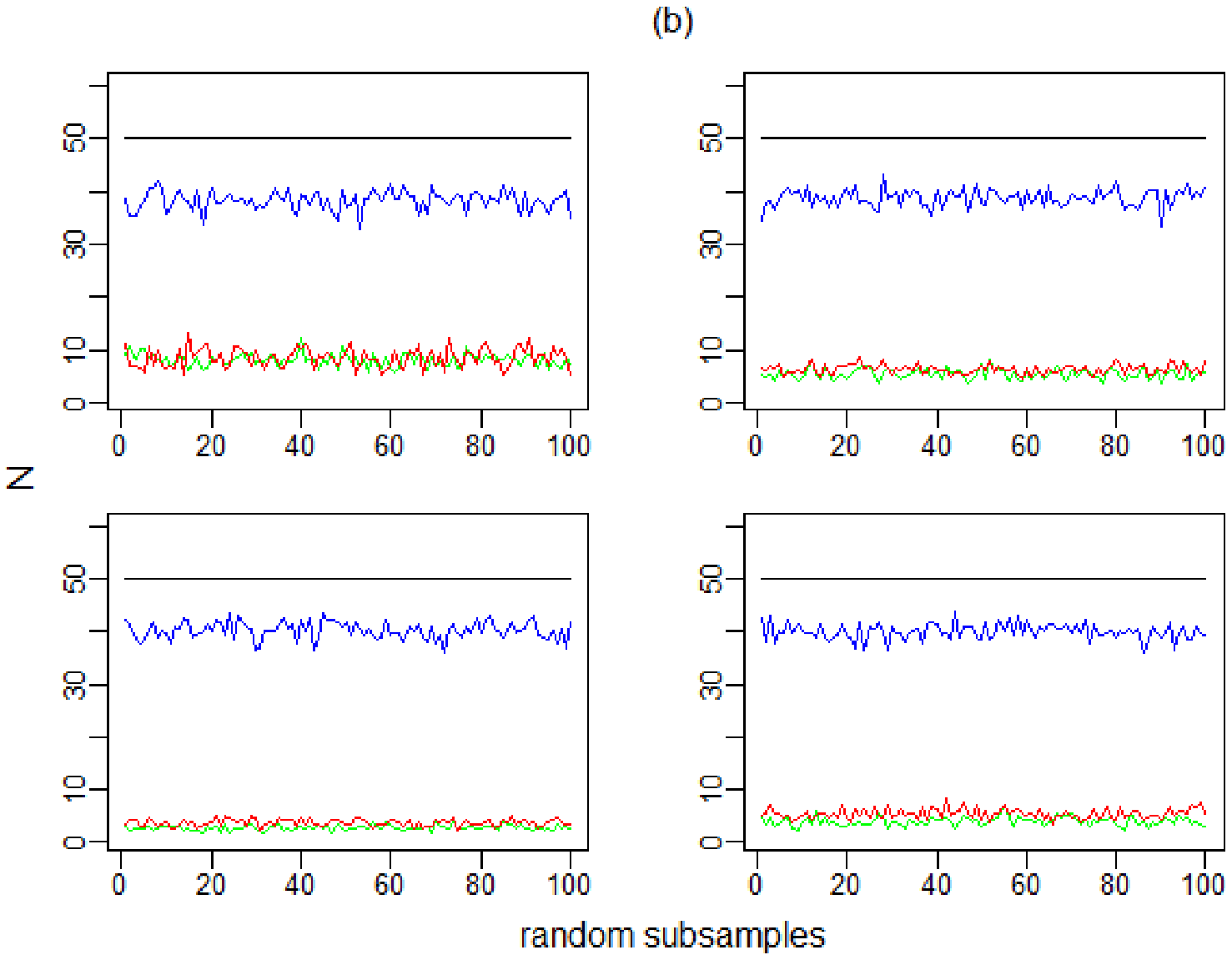}
\caption{\footnotesize{Effective sizes $\mathcal{N}$ for each of the 100 random subsamples. In panel (a), the size of the subsamples is 100, while in panel (b) is 50. The subsamples are of length $T = 3$ years with the plots covering together, inside each panel, the 2000-2012. Mesoscopic GMV portfolios in blue, RMT filtered in green, classical plug-in Markowitz in red.}}\label{8}
\end{center}
\end{figure}

In Figure \ref{8}, the effective size of the portfolios obtained with different allocation rules are displayed and compared: the compared allocation strategies are, again, the Markowitz GMV portfolios, the $1/N$ rule, the mesoscopic-based GMV portfolios and, finally, the noise-free GMV portfolios. It can be noticed that, independently from factors such as the time span, the subsample and the subsample size considered, the mesoscopic-based GMV portfolios are always much closer to the $1/N$ rule than those output by the classical Markowitz approach and the one based on RMT but filtering out only the random component. As it will be shown afterwards, this result brings non-trivial practical implications in terms of reliability.

\subsection{Mesoscopic community-based optimization scheme}

The adoption of the mesoscopic correlations leads to balanced portfolios closely tracking the equally-weighted investment plan. Let us now show how the portfolio optimization problem can be simply reformulated by taking into account the detected clusters of stocks, instead of the single ones, to further reduce the uncertainty characterizing each specific asset.

Let $N=N_{1}+N_{2}+\dots+N_{n}$ be the total number of asset, $n$ the number of detected communities, $N_{c}$ being the number of assets in a given community (denoted by the subscript $c \in \{1,2\dots n\}$). The problem, now, is that of finding the share of wealth $W_{c}$ which has to be invested into a given community, with $\omega_{c}=W_{c}/N_{c}$ being the share of wealth invested into the generic asset $i$ belonging to that community. The problem can be, thus, reformulated as follows

\begin{equation}
\begin{aligned}
\min_{\boldsymbol{\omega}} \quad & \boldsymbol{\omega'}\boldsymbol{\Sigma^{(g)}}\boldsymbol{\omega}\\
\textrm{s.t.} \quad & \mu_{p} = \boldsymbol{\omega'}\boldsymbol{\mu}\\
 & \sum_{c}^{n}W_{c}=1\\
 & \omega_{i} = \omega_{j} \;\;\;\; \forall \;\; i,j \in c.
\end{aligned}
\end{equation}
and is the same as the problem in 3 with the difference that weights are constrained to be equal for all the stocks belonging to the same community $c$; naturally, the total wealth share, being the sum of the wealth shares invested into each community, must still sum up to one. Reformulating the problem as in 16 leads to the minimization of the following objective function, where no constraint on the expected return is placed:

\begin{equation}
\sigma_{p}^{2} = \sum_{c=1}^{n}\omega_{c}^{2}\left[N_{c}\overline{\sigma}_{c}^{2} + N_{c}(N_{c}-1)\overline{\sigma}_{ijc}^{(g)}\right] + \sum_{c=1}^{n-1}\sum_{k=c+1}^{n}2\omega_{c}\omega_{k}\left[N_{c}N_{k}\overline{\sigma}_{ck}^{(g)}\right];
\end{equation}
notice that $\overline{\sigma}_{c}^{2}$ and $\overline{\sigma}_{ck}$ respectively denote the average of the variances inside a given community and the average of the mesoscopic covariances between assets belonging to different communities. The reliability analysis of the proposed approach will be assessed in the next section.

\begin{table}[t!]
\centering
\small
\begin{threeparttable}
\begin{tabular}{c c c c c c c}
\toprule
   &                     &         &  \multicolumn{2}{c}{Short-selling} & \multicolumn{2}{c}{No short-selling} \\                       
   
   & Time span & $\mathcal{R}_\textrm{equally}$ & $\mathcal{R}_\textrm{mesoscopic}$& $\mathcal{R}_\textrm{Markowitz}$ & $\mathcal{R}_\textrm{mesoscopic}$ & $\mathcal{R}_\textrm{Markowitz}$ \\

\cmidrule(lr){3-7}
 
 \multirow{3}*{$N=50$}  & $T_{1}$ &     0.53  &   \textbf{0.47}  & 0.79   &   0.48          & 0.68   \\ 
                        & $T_{2}$ &     4.06  &   \textbf{4.05}  & 5.67   & \textbf{4.05}    & 4.78     \\
                        & $T_{3}$ &     0.8   &   0.79      & \textbf{0.3}&   0.8           & \textbf{0.58}      \\
\cmidrule(lr){3-7} 
 \multirow{3}*{$N=100$} & $T_{1}$ &     0.17  &   \textbf{0.15}  & 1.12   &  0.23          & 0.53      \\ 
                        & $T_{2}$ &     1.38  &   2.07           & 3.74   &  \textbf{1.3}  & 2.72              \\
                        & $T_{3}$ & \textbf{0.27} & \textbf{0.27}& 0.31   &  0.4             & 0.3             \\
\cmidrule(lr){3-7}
 \multirow{3}*{$N=200$} & $T_{1}$ &     0.13 &   \textbf{0.11}   & 1.26   &   0.16         &  0.46    \\ 
                        & $T_{2}$ & \textbf{1.03}&   1.05        & 3.72   &   1.38         &  2     \\
                        & $T_{3}$ & \textbf{0.2}&   \textbf{0.2} & 0.43   &   0.26         &  \textbf{0.2} \\                    
\cmidrule(lr){3-7}
 \multirow{3}*{whole sample} & $T_{1}$ &     0.47  &  \textbf{0.45}   & 6.25   & \textbf{0.45}  &  0.61    \\ 
                        & $T_{2}$ &     4.75  &  \textbf{4.18}   & 17.11  &   4.88         &  6.65     \\
                        & $T_{3}$ &     0.8   &  \textbf{0.78}   & 2.32   &   0.79         &  \textbf{0.54}   \\   
\bottomrule
\end{tabular}
\caption{Reliability $\mathcal{R}$ for each strategy adopted and for each sample size, under different time spans, ranging from the $N=50$ case to the whole sample case ($N=450$). No randomization occured when the whole sample is taken. The measures  $\mathcal{R}_\textrm{mesoscopic}$ and $\mathcal{R}_\textrm{Markowitx}$ refer to the GMV portfolio cases with and without short-selling strategies. Entries in bold refer to the strategies with the best performances, both with and without short-selling constraints. Whenever the performance is worse than the equally weighted portfolio, the relative entry is not highlighted.}\label{T2}
\end{threeparttable}
\end{table}

\section{Reliability analysis}

Let us now compare the reliability $\mathcal{R}$ of the portfolios obtained by implementing the classical Markowitz portfolio optimization approach, the mesoscopic-based optimization approach, the mesoscopic plus community-based optimization approach and, finally, the heuristic equally-weighted strategy.

Both cases with and without short-selling will be analyzed, focusing on the GMV portfolios computed over different periods and for different sample sizes. Then, for the sake of completeness, we repeat the comparison considering 30 values of expected portfolio return $\mu_{p}$ to insert, as additional constraints, on the minimization problem. The different portfolios constituting the efficient frontier are, then, computed, and for each of them the index $\mathcal{R}$ is obtained, given the out-of-sample realized portfolios variances. The analyses are carried out over different time-spans and considering different sample sizes: the time-spans analyzed, i.e. $T_{1}=2000-2007$, $T_{2}=2004-2011$ and $T_{3}=2008-2015$ are divided in two additional subperiods of equal length by fixing $t_{0}$. Upon doing so, we create porfolios given the data collected over the period $t_{0}-\Delta t$ and quantify their out-of-sample performance over the period $t_{0}+\Delta t$. For what concerns the samples size, we randomly extract 100 subsamples out of the S\&P500 components, for each considered size.

\begin{table}[t!]
\centering
\small
\begin{threeparttable}
\begin{tabular}{c c c c c c}
\toprule
                               & Time span & $\mathcal{R}_\textrm{equally}$ & $\mathcal{R}_\textrm{mesoscopic}^{*}$ & $\mathcal{R}_\textrm{community}$ & $\mathcal{R}_\textrm{Markowitz}^{*}$ \\
\cmidrule(lr){2-6}
 \multirow{3}*{$\mathcal{R}$}  & $T_{1}$   &  0.47  &   0.45  & \textbf{0.41} & 0.61 \\ 
                               & $T_{2}$   &  4.75  &   4.88  & \textbf{3.45} & 6.65 \\
                               & $T_{3}$   &  0.8   &  0.78   & 0.74          & \textbf{0.54}  \\
\bottomrule
\end{tabular}
\caption{Comparison between the community-based portfolios and the other methodologies. With $\mathcal{R}_\textrm{mesoscopic}^{*}$ and $\mathcal{R}_\textrm{Markowitz}^{*}$ we indicate that we chose the best performance between the short-selling and no short-selling cases}\label{T3}
\end{threeparttable}
\end{table}

Average values computed for the $\mathcal{R}$ indices are reported in Table \ref{T2}. When short-selling is allowed and no constraint on the expected portfolio return is present, the Markowitz approach is always underperforming - the only exception being represented by the lowest-dimensionality case ($N=50$) - when compared with the $1/N$ and the mesoscopic-based optimization rule, irrespectively from the sample size and the time span considered. In addition, our methodology performs slightly better than the equally-weighted portfolio, thus revealing it to be the most reliable investment plan considered: the difference, however, is almost negligible, a result confirming the closeness of the mesoscopic-based optimization procedure and the equally-weighted strategy. From a purely mathematical perspective, imposing constraints is equivalent at letting a shrinkage operator act on the covariance matrix of the assets, an operation helping when the number of parameters to estimate is too large - and, as a consequence, estimation errors are large as well.

The poor performance of the GMV Markowitz portfolios is not a novel result, especially when no constraint about the possibility of exploiting short-selling strategies is imposed \citep[see][]{Frost,eichhorn1998using,britten1999sampling,jagannathan2003risk}. Our empirical analysis confirms these results: notice the huge improvement of the Markowitz approach compared to the situation in which short-selling was not allowed - although the mesoscopic-based optimization provieds better reliability indices in all cases except in period $T_3$.\\

Let us now check whether the clusters detected on $\mathbf{C}^{(g)}$ can be used as a further source of information. In particular, let us attach homogeneous optimal weights to stocks belonging to the same clusters, denoting with $\mathcal{R}_\textrm{community}$ the corresponding reliability index. To make the comparison as clear as possible, we compare the reliability of the community-based portfolios only with the best performing competing approach, in each time span. Results in Table \ref{T3} noticeably confirm the informativeness of the detected communities from a risk-management perspective. Portfolios in which optimal weights are recovered by constraining stocks in the same community to weigh the same further improve their reliability indices, outperforming both the equally-weighted and the mesoscopic-based strategies for all considered time spans. Still, Markowitz with the no short-selling constraint is the more reliable in $T_3$.

Let us now consider all approaches, i.e. the classical Markowitz one, the mesoscopic-based one and the mesoscopic plus community-based one and compute the reliability index for each portfolio of each efficient frontier. Results are summarized in Table \ref{T4}, where the $\mathcal{R}$ indices are ordered and compared between the different quartiles of the expected return distribution \footnote[8]{For each time span we take the historical expected return distribution of our assets (i.e. in-sample averages) and use the quartiles of the latter as input for the expected return constraints in the optimization problem.}: when adding constraints on expected returns, Markowitz outperforms our methodology, in time span $T_2$, when also the constraint on short-selling is imposed and in time span $T_3$ after the first quartile of the distributions? In time span $T_1$, instead, the community-based approach outperform the others. Overall, our approach is confirmed to perform better in all periods when we impose constraints only on expected returns but not on the weights. In particular, optimizing by taking into account the detected clusters stabilize the results, hence providing the best reliability.

Providing a deep explanation for such a result is hard given the higher degree of uncertainty introduced by the constraints on the expected returns. What is clear, however, is that cleaning the correlation matrices from both noise and systemic effects helps to ameliorate the reliability of the minimum variance portfolios and exploiting stocks communities identified through the mesoscopic correlation further improves the results. The same holds true when constraints on expected returns are imposed but allowing for short-selling strategies. When both constraints on returns and weights are in place, however, Markowitz approach is found to be hardly beatable.

\begin{table}
\centering
\small
\begin{threeparttable}
\begin{tabular}{c c c c c c c}
\toprule
    \textbf{Short-selling}    &      &   & & & & \\                     
\cmidrule (lr){2-7}                       
                       &                           & min.          & 1st quartile  & median        & mean          & 3rd quartile \\
\cmidrule(lr){2-7}
\multirow{3}*{$T_{1}$} & $\mathcal{R}_\textrm{community}$ & \textbf{0.34} & \textbf{0.39} & \textbf{0.43} & \textbf{0.5}  &  \textbf{0.52}  \\
                       & $\mathcal{R}_\textrm{mesoscopic}$      & 0.63          & 1.05          & 1.81          & 2.4           & 3.15 \\ 
                       & $\mathcal{R}_\textrm{Markowitz}$ & 1.13          & 1.46          &        2.1    & 2.5           & 3.31        \\
\cmidrule(lr){2-7}
\multirow{3}*{$T_{2}$} &  $\mathcal{R}_\textrm{community}$& \textbf{3.56} &  \textbf{3.96}& \textbf{4.09} & \textbf{4.02} & \textbf{4.19}  \\
                       & $\mathcal{R}_\textrm{mesoscopic}$      & 7.4           & 8.13          & 9.1           & 9.4           & 10.4        \\ 
                       & $\mathcal{R}_\textrm{Markowitz}$ & 4.05          & 7.07          & 7.67          & 7.85          & 8.5 \\
\cmidrule(lr){2-7}
\multirow{3}*{$T_{3}$} &  $\mathcal{R}_\textrm{community}$& 0.79          &  \textbf{0.81}& \textbf{0.81} & \textbf{0.81} &  \textbf{0.83}  \\     
                       & $\mathcal{R}_\textrm{mesoscopic}$      & \textbf{0.49} & 0.84          & 1.36          & 1.7           & 2.33          \\ 
                       & $\mathcal{R}_\textrm{Markowitz}$ & 0.88          & 1.07          & 1.4           &         1.51  & 1.85     \\
\cmidrule(lr){2-7}
     \textbf{No short-selling}        &       &  & & & &   \\ 
\cmidrule (lr){2-7}  
&                           & min.          & 1st quartile  & median        & mean          & 3rd quartile \\
\cmidrule(lr){2-7}
\multirow{3}*{$T_{1}$} & $\mathcal{R}_\textrm{community}$ & \textbf{0.006} & \textbf{0.14} & \textbf{0.20} & \textbf{0.26}  &  \textbf{0.46}  \\
                       & $\mathcal{R}_\textrm{mesoscopic}$      & 0.03   & 0.28    & 0.64   & 0.72     & 0.94    \\ 
                       & $\mathcal{R}_\textrm{Markowitz}$ & 0.02 & 0.22 & 0.48 & 0.48 & 0.75\\                       
\cmidrule (lr){2-7} 
\multirow{3}*{$T_{2}$} & $\mathcal{R}_\textrm{community}$ & 3.23 & 3.33 & 3.39 & 3.45  &  3.74  \\
                       & $\mathcal{R}_\textrm{mesoscopic}$      &     0.043   &  0.46  & 2.70   & 3.71      & 4.75   \\ 
                       & $\mathcal{R}_\textrm{Markowitz}$ &   \textbf{0.02}  & \textbf{0.42}   & \textbf{2.19}  & \textbf{1.52} & \textbf{2.37} \\
\cmidrule (lr){2-7} 
\multirow{3}*{$T_{3}$} & $\mathcal{R}_\textrm{community}$ & 0.74 & 0.76 & 0.77 & 0.76  &  0.78  \\
                       & $\mathcal{R}_\textrm{mesoscopic}$      & \textbf{0.03} & \textbf{0.31} & 0.56 & 1.34  & 2.64 \\
                       & $\mathcal{R}_\textrm{Markowitz}$ & 0.52 & 0.53 & \textbf{0.54} & \textbf{0.56}  & \textbf{0.59}    \\
\bottomrule                                                                              
\end{tabular}
\caption{Summary statistics of the reliability $\mathcal{R}$ indexes for the noise plus systemic free, classical Markowitz, and community-based efficient frontiers over different periods, with and without short selling strategies.}\label{T4}
\end{threeparttable}
\end{table}

\section{Discussion and conclusions}

In this work we investigated the mesoscopic structure of the stock market correlations that emerge after filtering out both microscopic (stock-specific noise) and macroscopic (market-wide trends) components. We showed that such mesoscopic correlations are the most stable over time, thereby encoding important information in the context of portfolio optimization. 
Indeed, we found that the noisy and the systemic components of the stock market are unstable, leading to biased and poor out-of-sample performances and being responsible for the surprising departure of the classical Markowitz investment prescription from the heuristic, equally weighted strategy. Upon filtering out these unstable components, the market can be partitioned into internally positively and mutually negatively correlated communities of stocks. We proposed to use these stable mesoscopic communities to construct portfolios characterized by higher levels of reliability in terms of predicted and realized risk. 

Results can be summarized as follows. The adoption of `noise- and systemic-free' correlations leads to an asset allocation which closely tracks, and slightly outperforms, the reliability of the heuristic equally weighted portfolio, while at the same time requiring a smaller number of assets over which the wealth need be effectively invested. In addition, both the equally weighted portfolios and the ones induced by the proposed optimization scheme have been found to be more reliable than the Markowitz plug-in estimator. Importantly, the reliability of portfolios can be further improved by performing the mesoscopic optimization while simultaneously accounting for the community to which a given stock belongs: remarkably, also when constraints on short-selling are imposed, this new methodology performs noticeably better than classical Markowitz.

Only when constraints on both weights and expected returns are imposed, the homogeneous community-based portfolios do not bring improvements compared to classical Markowitz - with the exception of the period $T_1=2000-2007$ and for few specific levels of targeted expected returns. Thus, the proposed methodology works well when focusing on the minimum-variance portfolio or when short-selling can be performed, suggesting the adoption of network clustering techniques for risk management applications. In particular, the uncovered mesoscale structure might bring insights about additional, and complementary, ways of creating stock market indices to monitor market trends - something which might be the object of further studies aimed at understanding co-movements between industries and sectors in the stock market.

\bibliographystyle{chicago}
\bibliography{mybib}

\appendices

\section{Brief analytical clarifications with the 2-asset case}
Consider an investor who splits her wealth between $N = 2$ assets and want to minimize the variance of her investment. The problem to solve simply is
\begin{equation}
\min_{\boldsymbol{\omega_{1}}} \;\; \omega_{1}^{2}\sigma_{1}^{2} + (1-\omega_{1})^{2}\sigma_{2}^{2} + 2\omega_{1}(1-\omega_{1})C_{12}\sigma_{1}\sigma_{2},
\end{equation}
whose first order condition is 
\begin{equation}
2\omega_{1}\sigma_{1}^{2} - 2(1-\omega_{1})\sigma_{2}^{2} + 2(1 - 2\omega_{1})C_{12}\sigma_{1}\sigma_{2} = 0
\end{equation}
implying the following optimal wealth allocation with respect to asset 1 
\begin{equation}
\omega_{1}^{*} = \frac{\sigma_{2}^{2} - C_{12}\sigma_{1}\sigma_{2}}{\sigma_{1}^{2} + \sigma_{2}^{2} - 2C_{12}\sigma_{1}\sigma_{2}}.
\end{equation}
Given the decomposition in (8), we know that the noise-free correlation coefficients and covariances are $C_{ij}=C_{ij}^{(g)} + C_{ij}^{(m)}$ and $\sigma_{ij} = \sigma_{ij}^{(g)} + \sigma_{ij}^{(m)}$, we thus write
\begin{equation}
\omega_{1}^{*} = \frac{\sigma_{2}^{2} - (\sigma_{12}^{(m)}+\sigma_{12}^{(g)})}{\sigma_{1}^{2} + \sigma_{2}^{2} - 2(\sigma_{12}^{(m)}+\sigma_{12}^{(g)})}.
\end{equation} 
For a risk minimizer investor who filters out the systemic induced covariances being aware of its temporarily nature, or equivalently in absence of significant systemic comovements, the optimal adjusted weight is the one obtained using the mesoscopic covariances
\begin{equation}
\omega_{1}^{adj} = \frac{\sigma_{2}^{2} - C_{12}^{(g)}\sigma_{1}\sigma_{2}}{\sigma_{1}^{2} + \sigma_{2}^{2} - 2C_{12}^{(g)}\sigma_{1}\sigma_{2}}.
\end{equation} 

This difference can be easily quantified taking $\Delta\omega_{1}^{*}=\omega_{1}^{*} - \omega_{1}^{adj}$, which after some manipulation and terms  rearranging yields
\begin{equation}
\Delta\omega_{1}^{*} = \frac{2\sigma_{12}^{(m)}\sigma_{2}^{2} \;-\; \sigma_{12}^{(m)}(\sigma_{1}^{2}+\sigma_{2}^{2})}{(\sigma_{1}^{2}+\sigma_{2}^{2})^{2} - 4\sigma_{12}^{(g)}(\sigma_{1}^{2}+\sigma_{2}^{2}+\sigma_{12}^{(m)}+\sigma_{12}^{(g)})}.
\end{equation}
If $\sigma_{12}^{(m)} = 0 \rightarrow \Delta\omega_{1}^{*}=0$ and no difference in the wealth allocation occurs.\\
Otherwise, $\sigma_{12}^{(m)}>0 \rightarrow$ $\Delta\omega_{1}^{*}>0$ if $\sigma_{2}^{2}>\sigma_{1}^{2}$, which clarify the rebalancing of the portfolio stated in the paper and empirically displayed.

\section{Technical steps GMV decomposition}
Consider the solution of the GMV portfolio 
\begin{equation}
w_{gmv} = \frac{\boldsymbol{\Sigma}^{-1}\boldsymbol{1}}{{\boldsymbol{1}}^{t}\boldsymbol{\Sigma}^{-1}\boldsymbol{1}},
\end{equation}
and the spectral decomposition of the covariance matrix 
\begin{equation}
\boldsymbol{\Sigma}^{-1} =\boldsymbol{PD^{-1}P^{-1}}.
\end{equation}
where $\boldsymbol{D}$ is the diagonal matrix from which we are able to identify the eigenvalues associated to random covariances exploiting the \textit{MP-Law}, and the biggest one associated to the systemic component. Thus, $\boldsymbol{D}$ can be splitted as 
\begin{equation}
\boldsymbol{D} = \boldsymbol{D^{(r)} + D^{(g)} + D^{(m)}}
\end{equation}
and its inverse can be obtained by simply replacing each non zero element in the main diagonal (i.e. eigenvalues) with its reciprocal, having 
\begin{equation}
\boldsymbol{D}^{-1} = \boldsymbol{D_{(r)}^{-1} + D_{(g)}^{-1} + D_{(m)}^{-1}}.
\end{equation}
Combining the above equations we get
\begin{align*}
\boldsymbol{\Sigma^{-1}} &= \boldsymbol{PD_{r}^{-1}P^{-1} + PD_{g}^{-1}P^{-1} + PD_{m}^{-1}P^{-1}}\\
\\
&=\boldsymbol{ \Sigma_{r}^{-1} + \Sigma_{g}^{-1} + \Sigma_{m}^{-1}}
\end{align*}
which allows to split the GMV solution  as
\begin{align*}
w_{gmv} &= \frac{\Sigma_{r}^{-1}\boldsymbol{1}}{\boldsymbol{1}^{t}\Sigma^{-1}\boldsymbol{1}} + \frac{\Sigma_{g}^{-1}\boldsymbol{1}}{\boldsymbol{1}^{t}\Sigma^{-1}\boldsymbol{1}} + \frac{\Sigma_{m}^{-1}\boldsymbol{1}}{\boldsymbol{1}^{t}\Sigma^{-1}\boldsymbol{1}}\\
\\
        &= w_{gmv}^{(r)} + w_{gmv}^{(g)} + w_{gmv}^{(m)}. 
\end{align*}

\section{Details on the implementation of the community-based optimization procedure}
Consider the variance of the portfolio
\begin{equation}
\sigma_{p}^{2} = \sum_{i=1}^N \omega_{i}^2\sigma_{i}^2 + \sum_{i=1}^{N}\sum_{i \neq j}\omega_{i}\omega_{j}\sigma_{ij}.
\end{equation}
Remember that $N = N_{1} + N_{2} + ... + N_{n}$ is the total number of asset, $n$ the number of detected communities, and $N_{c}$ the number of assets in a given community denoted by the subscript $c \in \{1,2, ... n\}$. We also drop the superscript $(g)$ taking for granted that we always refer to the  covariance between assets already filtered from both noise and systemic effects. Maximizing with respect to the $n$ detected communities, so to have homogeneous weights inside a given community, can be achieved by splitting the variance of the portfolio as follows
\begin{equation}
\begin{aligned}
\sigma_{p}^{2} \quad &= \sum_{i=1}^{N_{1}}\omega_{1}^2\sigma_{i1}^2 + \sum_{i=1}^{N_{2}}\omega_{2}^2\sigma_{i2}^2 + \dots + \sum_{i=1}^{N_{n}}\omega_{n}^2\sigma_{in}^2 \\
  \quad & + \sum_{i=1}^{N_{1}}\sum_{i \neq j}\omega_{1}^{2}\sigma_{ij1} +  \sum_{i=1}^{N_{2}}\sum_{i \neq j}\omega_{2}^{2}\sigma_{ij2} + \dots + \sum_{i=1}^{N_{n}}\sum_{i \neq j}\omega_{n}^{2}\sigma_{ijn} \\
  \quad & + \sum_{i=1}^{N_{1}}\sum_{j=1}^{N_{2}}\omega_{1}\omega_{2}\sigma_{ij12} + \sum_{i=1}^{N_{1}}\sum_{j=1}^{N_{3}}\omega_{1}\omega_{3}\sigma_{ij13} + \dots + \sum_{i=1}^{N_{1}}\sum_{j=1}^{N_{n}}\omega_{1}\omega_{n}\sigma_{ij1n} \\
  \quad & + \sum_{i=1}^{N_{2}}\sum_{j=1}^{N_{3}}\omega_{2}\omega_{3}\sigma_{ij23} + \dots + \sum_{i=1}^{N_{2}}\sum_{j=1}^{N_{n}}\omega_{2}\omega_{n}\sigma_{ij2n} \\    
  \quad & \vdots \\
  \quad & +  \sum_{i=1}^{N_{n-1}}\sum_{j=1}^{N_{n}}\omega_{n-1}\omega_{n}\sigma_{ij(n-1)n}\\
\end{aligned}
\end{equation}
which is equivalent to
\begin{equation}
\sigma_{p}^{2} = \sum_{c=1}^{n}\omega_{c}^{2}N_{c}\bar{\sigma}_{c}^{2} + \sum_{c=1}^{n}\omega_{c}^{2}N_{c}(N_{c}-1)\bar{\sigma}_{ijc} + \sum_{c=1}^{n-1}\sum_{k=c+1}^{n}2\omega_{c}\omega_{k}N_{c}N_{k}\bar{\sigma}_{ck}
\end{equation}
Thus the objective function to minimize with respect to the community weights become
\begin{equation}
\sigma_{p}^{2} = \sum_{c=1}^{n}\omega_{c}^{2}\left[N_{c}\bar{\sigma}_{c}^{2} + N_{c}(N_{c}-1)\bar{\sigma}_{ijc}\right] + \sum_{c=1}^{n-1}\sum_{k=c+1}^{n}2\omega_{c}\omega_{k}\left[N_{c}N_{k}\bar{\sigma}_{ck}\right]
\end{equation}
with $W_{c} = \omega_{c}N_{c}$ being the total share of wealth invested in community $c$.

\end{document}